\documentclass[article,nofootinbib,groupaddress,floatfix,letterpaper,preprint,showpacs,showkeys]{revtex4}
\usepackage{graphicx}
\usepackage{amsmath}
\usepackage{dcolumn}
\usepackage{subfigure}
\usepackage{multirow}
\usepackage{mathtools}
\usepackage{epstopdf}
\usepackage{mathptmx}
\providecommand{\keywords}[1]{\textbf{\textit{Index terms---}} #1}

\begin{document}

\title{On-Chip Cavity Optomechanical Coupling}

\author{Bradley D Hauer, Paul H Kim, Callum Doolin, Allison JR MacDonald, Hugh Ramp and John P Davis\footnote{Corresponding author: jdavs@ualberta.ca}}
\affiliation{Department of Physics, University of Alberta, T6G 2E1 Edmonton, AB, Canada}

\begin{abstract}

On-chip cavity optomechanics, in which strong co-localization of light and mechanical motion is engineered, relies on efficient coupling of light both into and out of the on-chip optical resonator.  Here we detail our particular style of tapered and dimpled optical fibers, pioneered by the Painter group at Caltech, which are a versatile and reliable solution to efficient on-chip coupling.  First, a brief overview of tapered, single mode fibers is presented, in which the single mode cutoff diameter is highlighted. The apparatus used to create a dimpled tapered fiber is then described, followed by a comprehensive account of the procedure by which a dimpled tapered fiber is produced and mounted in our system. The custom-built optical access vacuum chambers in which our on-chip optomechanical measurements are performed are then discussed.  Finally, the process by which our optomechanical devices are fabricated and the method by which we explore their optical and mechanical properties is explained.  It is our expectation that this manuscript will enable the novice to develop advanced optomechanical experiments.  

\end{abstract}



\maketitle

\section*{Background}

State-of-the-art nanofabrication technologies have allowed for a drastic reduction in the size, and increase in quality, of nanomechanical systems, which have been the driving force behind radically increasing the sensitivity of numerous devices. Examples include accelerometers \cite{krause}, mass sensors \cite{schmid,yang,jensen,chaste,li2}, electrometers \cite{cleland}, temperature sensors \cite{larsen,zhang}, force transducers \cite{gavartin,moser} and biosensors \cite{fritz,mertens,ndieyira,gupta}. The mass of a sensor and its ability to precisely measure physical quantities are intimately related, with smaller devices having superior sensitivity \cite{chaste,li2,moser}. However, as we continue to reduce device volume, it is difficult to find transduction methods that scale appropriately. Furthermore, it is generally the case that the target quantity is measured through the nanomechanical device's motion, hence as we move to more sensitive devices, we require a detection method with comparable precision. A solution to these issues has been found in the field of cavity optomechanics, which allows for quantum-limited, sub-am/$\sqrt{\rm Hz}$ displacement sensitivity \cite{arcizet,schliesser} and device masses down to the pico/femto-gram range \cite{doolin,stapfner,eichenfield}.

Optomechanics describes the coupling of the mechanical motion of a device to an optical field, often to manipulate or detect its motion.  It is advantageous to use an optical cavity, such as a whispering gallery mode (WGM) resonator, to provide this field, as the light in the optical cavity is able to sample the mechanics many times due to its long photon lifetime and leads to resonantly enhanced optomechanical coupling.  In such a system, the motion of the mechanical device shifts the resonance frequency and phase of the optical cavity. By detecting this signal, it is possible to infer the motion of the device. At the same time, photons inside the cavity apply a radiation pressure to the mechanical resonator \cite{hossein}, which can be used to optomechanically dampen or amplify its motion, lending to a large number of interesting phenomena \cite{kippenberg}. Cavity optomechanical systems have been realized in a number of different geometries, including photonic crystal cavities \cite{eichenfield}, Fabry-P\'erot etalons \cite{jayich,favero}, WGM resonators \cite{anetsberger,park,kim} and electronic microwave cavities \cite{teufel}.

It is advantageous to fabricate cavity optomechanical devices on-chip, as it is therefore possible engineer both the optical and mechanical resonators to certain desired specifications. Modern nanofabrication technologies make it possible to produce devices with extremely accurate predefined dimensions, allowing feature sizes as small as 100 nm for foundry-based deep ultraviolet (DUV) optical lithography \cite{selvaraja} and 2 nm with electron beam lithography \cite{manfrinato}. This enables precise tailoring of important device parameters, such as the gap between mechanical and optical resonators, which controls the optomechanical coupling in our devices \cite{kim}. Furthermore, devices fabricated using top-down lithography can be integrated into electronic on-chip devices \cite{winger} and, in the case of optical lithography, can easily be mass produced.

However, difficulties arise when trying to couple light into these on-chip devices, as a high on-chip density and planar geometry require a precise optical probe which can couple exclusively to a particular device. This problem has been solved by using dimpled tapered fibers \cite{kim,michael}, providing efficient and maneuverable coupling to on-chip optomechanical devices. 

In this article, we describe a process by which such a coupling system is produced, outlining the necessary steps while assuming no special knowledge \textit{a priori}. We begin by investigating the fundamentals of tapered optical fibers, as well as describing an apparatus which can be used to fabricate and dimple them. Following this is a discussion of the procedure by which dimpled tapered fibers are produced. Custom-built optical access vacuum chambers, which are used for optomechanical coupling to on-chip devices, are also detailed. Finally, we explain our method of coupling to optomechanical devices with tapered fibers. Using these systems, we have been able to demonstrate the first ever on-chip optomechanical torsional sensors \cite{kim}, as well as multidimensional detection of high frequency microcantilevers \cite{doolin} suitable for force sensing applications.

\section*{Discussion}
\subsection*{Single Mode Tapered Optical Fibers}

A crucial element in any optomechanical device is the method by which the optical field is injected, and subsequently collected, from the optical resonator in the system. While a number of different options exist, including free space optical coupling \cite{fiore}, grating couplers \cite{li} and fiber-to-waveguide coupling \cite{groblacher,cohen}, we have chosen to use direct coupling from tapered optical fibers \cite{eichenfield,knight,cai,barclay,srinivasan2}. Tapered fibers are more efficient, and require less on-chip space, than grating couplers, while free-space coupling is inconsistent with on-chip devices.  It may prove that fiber-to-waveguide coupling \cite{cohen} is more efficient and stable than tapered fibers, but the versatiliy and maneuverability of tapered fibers remains a significant advantage.

\begin{figure}[h!]
\includegraphics[width = \columnwidth]{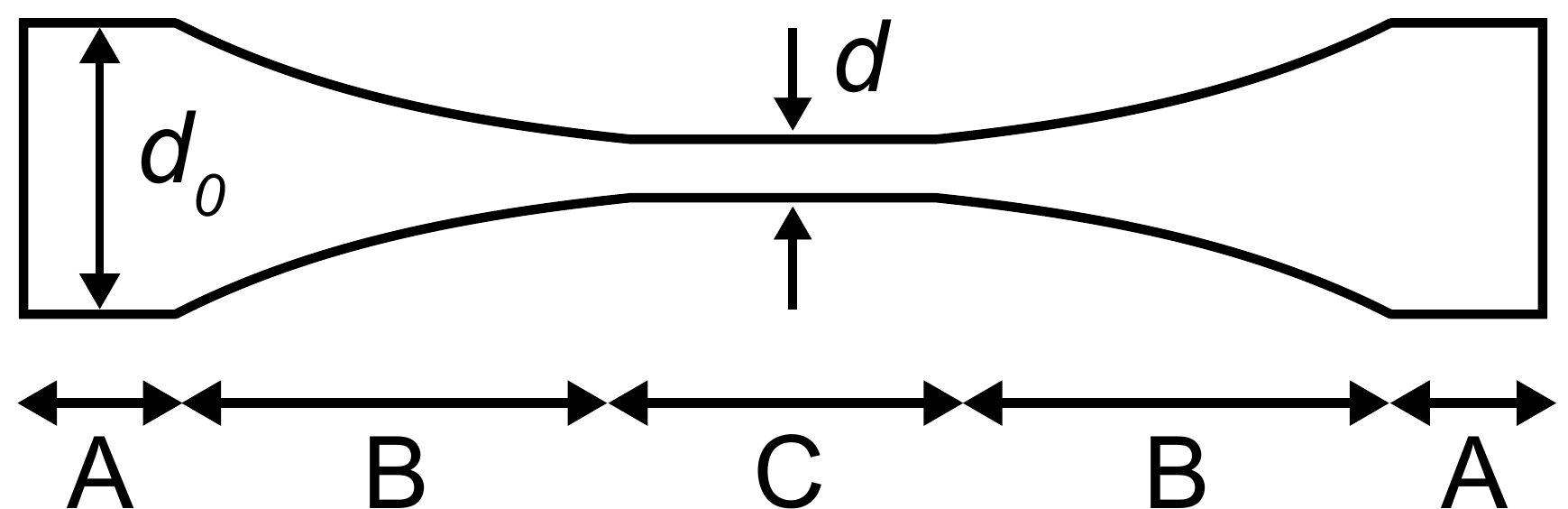}
\caption{{\bf Tapered Fiber Regions:} Diagram of a tapered fiber. Region A is the {\it unstretched fiber}, where the fiber remains the original unperturbed diameter $d_0$. Region B is the {\it taper transition} over which the unstretched diameter is adiabatically reduced to the minimum diameter $d$ maintained throughout Region C, which known as the {\it taper waist}. A constant hot-zone $L$ is assumed such that the taper transition has an exponential profile.}
\label{fibdia}
\end{figure}

A tapered fiber is a standard optical fiber (silica core surrounded by a higher index cladding) that has had its initial diameter adiabatically reduced over a small length known as the {\it tapered region}.  This can be performed either through hydrofluoric acid etching of an optical fiber \cite{zhang2,laine}, or by the heat-and-pull method \cite{brambilla,ward,ding,tong}. In this latter method, a small region of an optical fiber, known as the {\it hot-zone}, is heated to the point of melting and subsequently stretched to reduce its diameter. The final tapered fiber will then consist of three regions, the initial unstreched fiber, the taper transition, and the taper waist, all of which are detailed in Fig.~\ref{fibdia}.  From a conservation of mass argument, it can be shown that for a constant hot-zone of length $L$, which is produced in the case of a stationary flame, the taper transition is exponential \cite{birks}. Using a constant pull speed $v$, this results in a taper waist diameter $d$ that decreases with pull time $t$ according to

\begin{equation}
d = d_0 e^{-v t / L},
\label{expd}
\end{equation}

\noindent where $d_0$ is the diameter of the initial untapered fiber.

Following the heat-and-pull process, a new air-clad core exists in the taper waist, comprised of a composite material with an effective index determined by the indices and relative sizes of the initial core and cladding. This region can be modeled as a long, dielectric cylinder, for which Maxwell's equations can be solved analytically to determine the electromagnetic modes of the core (cladding) in terms of Bessel (modified Bessel) functions of the first (second) kind, as described in \cite{arnaud}. In general, such a structure will support many modes, lending to the description of a multimode fiber. However, once the fiber's diameter drops below a critical value, known as the {\it single mode cut-off diameter}, only a single guided mode remains in the fiber, labeled the hybrid $HE_{11}$ mode \cite{borselli}, as all other spatial modes decay evanescently. We look to determine this critical diameter for light with a free space wavelength, $\lambda$, traveling in a fiber with a core of index of refraction, $n_{co}$, surrounded by a cladding with index, $n_{cl}$. This is done by matching the electromagnetic fields within the core and cladding according to the boundary conditions given by Maxwell's equations \cite{jackson}, resulting in the following expression for the single remaining mode

\begin{equation}
\left[ \frac{J_1(x)}{x J_0(x)} + \frac{K_1(y)}{y K_0(y)} \right]\left[ \frac{n_{co}^2}{n_{cl}^2} \frac{J_1(x)}{x J_0(x)} + \frac{K_1(y)}{y K_0(y)} \right] = 0.
\label{xyrel}
\end{equation}

\noindent In the above equation, $J_{\nu}(x)$ is the Bessel function of the first kind and $K_{\nu}(x)$ is the modified Bessel function of the second kind. As well, $x = \frac{d }{2}\sqrt{k_{co}^2 - \beta^2}$ and $y = \frac{d}{2} \sqrt{\beta^2 - k_{cl}^2}$, where $k_{co}=2 \pi n_{co} / \lambda$ and $k_{cl} = 2 \pi n_{cl} / \lambda$ are the magnitudes of the wavevector in the core and cladding, respectively, and $\beta$ is the fiber's propagation constant. From these definitions, we can immediately derive the expression 

\begin{equation}
x^2 + y^2 = \frac{\pi^2 d^2 (n_{co}^2 - n_{cl}^2)}{\lambda^2}.
\label{xysq}
\end{equation}

\noindent Using this relationship between $x$ and $y$, we are able to determine the fiber diameter $d_c$ such that Eq.~\ref{xyrel} has only one solution for $d <d_c$, indicating the point at which the penultimate mode ceases to exist. This value is the single mode cut-off diameter and is determined by numerically calculating the solutions to Eq.~\ref{xyrel} while iteratively increasing $d$ until a second solution emerges.

It is also possible to determine an analytic expression for $d_c$ in the weakly-guiding approximation (WGA) \cite{gloge}. In this case, we take $n_{co} \approx n_{cl}$, so that Eq.~\ref{xyrel} becomes

\begin{equation}
\frac{x J_0(x)}{J_1(x)} = - \frac{y K_0(y)}{K_1(y)}.
\label{xyrelapprox}
\end{equation}

\noindent For the single mode cut-off, $y=0$ ({\it i.e.} $\beta = \pm k_{cl}$), which is physically interpreted as the mode evanescently decaying into the cladding. Using $\displaystyle \lim_{y \to 0} \frac{y K_0(y)}{K_1(y)} = 0$, we see that Eq.~\ref{xyrelapprox} has solutions when $xJ_0(x) = 0$. One solution will always exist for $x=0$, corresponding to the single remaining mode below cut-off. The penultimate mode comes into existence when $J_0(x)=0$ for the first time, which occurs at $x = 2.4048$. Therefore, we can find $d_c$ by rearranging Eq.~\ref{xysq} to get \cite{borselli}

\begin{equation}
d_c = \frac{ 2.4048 \lambda}{\pi \sqrt{n_{co}^2 - n_{cl}^2}}.
\label{cutoffd}
\end{equation}

\noindent The validity of the WGA is confirmed by comparing the results of Eq.~\ref{cutoffd} to numerically calculated cutoff diameters, which are summarized for a number of situations in Table \ref{cutofftab}.

At these diameters there exists a significant evanescent field surrounding the waist region of the tapered fiber. This allows for substantial overlap between an optical resonator's modes and the fiber's guided light when it is approached to an optical cavity. Likewise, light trapped inside the cavity will couple back into the fiber, which will be carried away as optomechanical signal.

While this type of straight tapered fiber is useful for coupling to a single off-chip device, such as a microsphere \cite{knight}, it is difficult to use as probe of on-chip devices, although it can be done if the device is cleaved to hang over the edge of the chip \cite{anetsberger} or isolated using a mesa \cite{srinivasan}.  Instead, it is useful to introduce a small dimpled region to the fiber, producing a portion of the taper waist that can be used as a probe of a single on-chip optomechanical device \cite{michael}.  
   When combined with a precise positioning system, this allows for sampling of numerous devices with the localized coupling region at the tip of the dimple of the tapered fiber. 

\subsection*{Tapered Fiber Puller}

\begin{figure}[h!]
\includegraphics[width = \columnwidth]{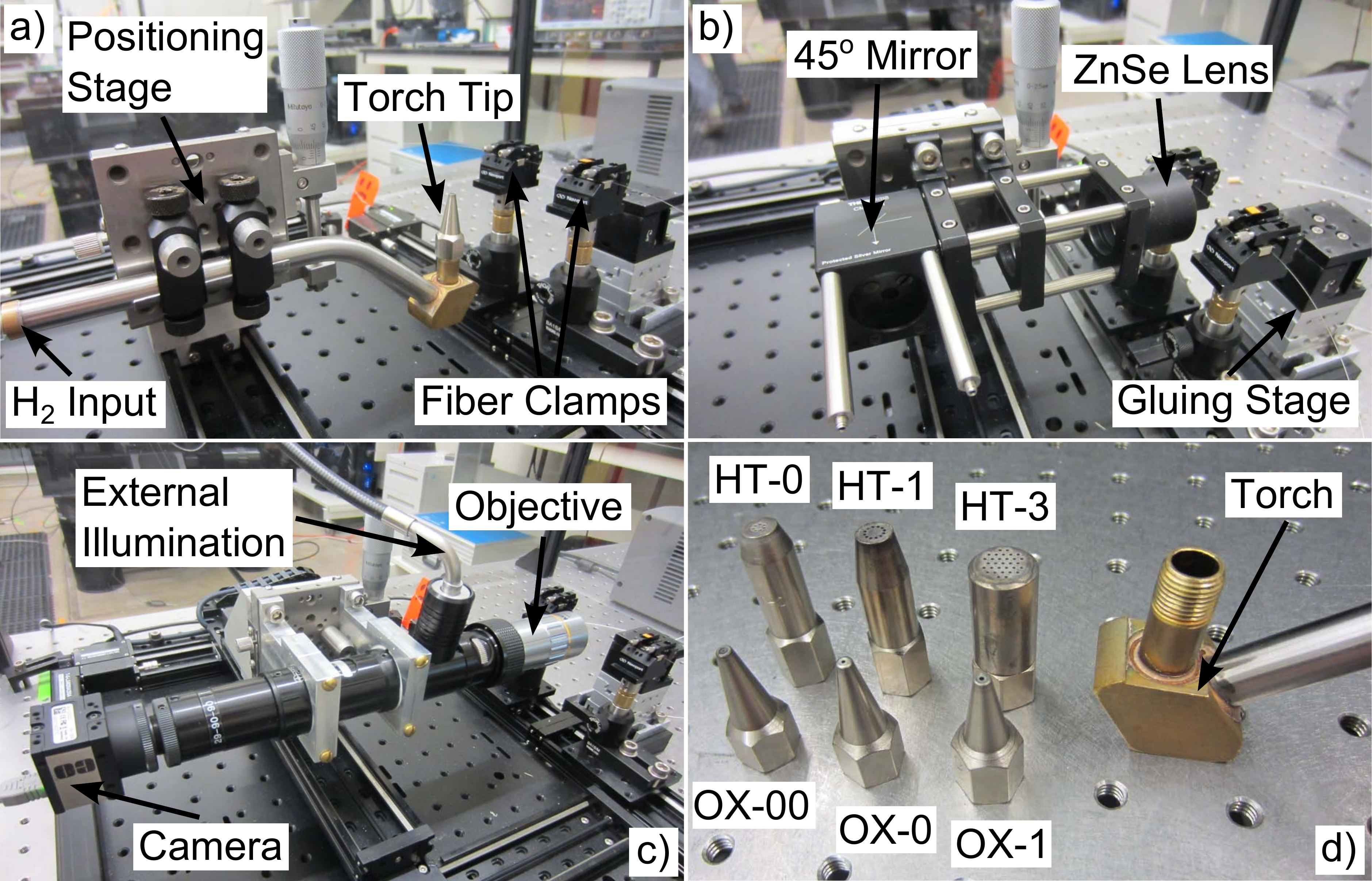}
\caption{{\bf Tapered Fiber Puller:} Pictures of the tapered fiber pulling apparatus with a) the hydrogen torch, b) the cage mounted ZnSe lens and c) the microscope imaging system mounted on the positioning gantry. d) Picture of the hydrogen torch along with the HT and OX series torch tips.}
\label{tappull}
\end{figure}

To produce tapered fibers, we use a heat-and-pull method in which a flame from a hydrogen torch is used to soften or melt an optical fiber while simultaneously stretching it at a constant speed. In our system, we produce this flame using a custom-built mountable hydrogen torch, as seen in Fig.~\ref{tappull}a, which is threaded using a 7/16"-24 die (McMaster-Carr, Part No. 26005A128) producing standard threads that allow for interchangeability of torch tips. The tips we use are the HT and OX series purchased from National Torch (see Fig.~\ref{tappull}d), which provide a wide variety of flame sizes useful for producing different sizes of tapered fibers. The hydrogen torch is fed by a needle valve-controlled line, allowing for a very small and stable flame using the OX-00 torch tip, with a single 0.51 mm diameter hole. This tip is chosen because it produces compact tapers (less than 1 cm in total length) which are ideal for our fiber holders, while maintaining a relatively high transmission efficiency (up to $\sim$80\%).  

The hydrogen torch is mounted on a three-axis positioning system, consisting of automated $xy$-translation in the plane of the optical table on which the apparatus is mounted, along with perpendicularly oriented manual $z$-adjustment. The $xy$-translation system is based on a Zaber T-G-LSM200A200A two-axis gantry system.  Each orthogonal axis is driven by a Zaber T-LSM200A linear motorized stage, allowing for a total travel range of 200 mm in either dimension with a minimum step size of 50 nm. Manual $z$-adjustment is provided by a New Focus 9063-COM gothic-arch translation stage mounted using a New Focus 9063-A angle bracket. The stage is manipulated by a Mitutoyo No. 906912 micrometer, providing a 25 mm travel range with 10 $\mu$m resolution. This system is used for precise and reproducible placement of the hydrogen torch flame as it heats the fiber, which is an important element required to consistently produce high quality tapered fibers. The fiber itself is held using two Newport 466A-710 dual arm V-groove fiber holders, each of which is connected to an adjustable optical post mounted on a Zaber T-LSM100A linear motorized stage. Each stage has a travel range of 100 mm with a resolution of 50 nm and can pull the melted fiber at speeds up to 7 mm/s. All of the Zaber stages are automated in software, allowing for precise, reproducible $xy$-positioning of the torch gantry, as well as the ability to set a consistent pull speed. The adjustable optical posts help to ensure that the fiber is level, as proper alignment is crucial for producing a low-loss taper. This entire setup is surrounded by a protective box, built from optical rails and acrylic sheets, which helps reduce flame instability due to air currents, as well as preventing contaminants from entering the system. 

Another method by which a fiber can be tapered is using a CO$_2$ laser, which produces radiation with a wavelength ranging from 10.2 - 10.8 $\mu$m \cite{ward}. Absorption of these photons by an optical fiber causes it to heat in proportion to the intensity of the beam and the cross section of the fiber being irradiated. Therefore, the power of a CO$_2$ laser must be carefully controlled while pulling a fiber in order to ensure even heating. CO$_2$ lasers have also been used as a heat source for other processes, namely in the production of high-$Q$ silica WGM resonators, such as microspheres \cite{collot} and bottles \cite{pollinger}.

We integrate a CO$_2$ laser into our fiber pulling system by replacing our hydrogen torch with a cage mount system (see Fig.~\ref{tappull}b) containing a 45 degree cube-mounted silvered mirror (Thorlabs - Product No. CM1-P01) and a plano-convex ZnSe lens (Thorlabs - Product No. LA7542-F) with a focal length of 25.4 mm, which can be used to focus the intense, infrared radiation from the laser onto the fiber. Attaching our lens to the torch positioning system, we gain full control of its position. This allows for defocusing of the CO$_2$ laser beam, effectively controlling both the size and temperature of the hotspot on the fiber. Furthermore, the focused beam can be scanned along the fiber, allowing for a movable hotspot, which is required for bottle fabrication \cite{pollinger}. Finally, manual height adjustment of the cage mount focusing system allows us to ensure that the CO$_2$ beam will hit the center of the lens, reducing aberration. 

It is also possible to attach a microscope imaging system directly to our torch positioning gantry, as shown in Fig.~\ref{tappull}c. The microscope is comprised of a 10X M Plan Apo long working distance infinity-corrected objective (Edmund Optics - Stock No. \#59-877) attached to an Optem Zoom 70XL lens system, allowing for 70$\times$ magnification of the setup. This image is recorded using an Edmund Optics EO-5012C color USB webcam, providing a video feed to a nearby computer. To ensure proper lighting and image quality, light from an external Metaphase MP-LED-150 microscope LED illuminator is coupled into the lens system's coaxial illumination port using a fiber optic waveguide (Edmund Optics - Stock No. \#39-368). This system is very useful, as it allows for real time imaging of our completed tapered fibers (and other fabricated optical components) when dimpling or attaching it to its holder, with full three-axis control. 

As tapered fibers are quite fragile, it is difficult to move them without breaking. For this reason, we first attach the tapered fiber to a holder, creating a more robust system which can easily be relocated. To this end, we have also included a manually adjusted Newport Compact Dovetail DS40-XYZ three-dimensional linear positioning stage in our system, which allows for 1 $\mu$m sensitivity over a travel range of 14 mm in each of $x$ and $y$ and 5 mm in $z$. This stage allows us to properly position and align the fiber holder, as well as gradually approach it to the fiber for gluing. In addition, it is used to position the fiber mold used in the dimpling process, which must be approached and raised precisely at the thinnest point of the tapered fiber.

Another important aspect of the tapered fiber puller is the fiber transmission monitoring system, which allows us to determine the point at which the taper becomes single mode, as well as assess fiber losses due to tapering. To do this, we measure the transmitted power of light from a New Focus Velocity 6330 tunable diode laser through the fiber during the pulling process. To control the amount of injected power, laser light is first passed through a Thorlabs VOA50-APC variable optical attenuator (VOA) before it is coupled into the fiber using a mechanical splicer (Fiber Instrument Sales elastomeric lab splice - Part No. FIS114012), in which two straight cleaved fiber ends are butt coupled to each other with the aid of index matching gel (Fiber Instrument Sales matching gel - Part No. F10001V). Likewise, the fiber is mechanically spliced on its opposite end to a patch cable connected to a New Focus Model 1811 IR DC-125 MHz low noise photoreceiver. The DC signal from this photodiode is split off and recorded using an NI USB-6259 BNC DAQ card for the duration of a fiber pull, providing a record of transmission vs pull time, as seen in Fig.~\ref{fibtrans}.

\subsection*{Fiber Tapering Procedure}

To create tapered fibers, we begin with a Corning SMF-28e optical fiber that has a silica core and cladding diameter of 8.2 $\mu$m and 125 $\mu$m, respectvely, all of which is protected by an acrylate coating which extends out to a diameter of 245 $\mu$m. The indices of refraction and dimensions of the core and cladding are chosen such that this original fiber is single mode for wavelengths exceeding 1260 nm, which includes both the dispersionless and minimum loss wavelengths in silica of 1310 nm and 1550 nm, respectively. 

To begin the tapering process, the acrylate coating is removed using a Micro-Strip$^{\textregistered}$ stripping tool over a region approximately 3 cm long in the center of an SMF-28e fiber around one meter in total length. This section of stripped fiber is subsequently cleaned using a solvent to remove any remaining acrylate. The tapering occurs in this stripped region, where the flammable acrylate has been removed. In addition, the two ends of the fiber are stripped of acrylate and cleaved flat using an Ericsson EFC11 fiber cleaver. Utilizing the mechanical splicers and index matching gel described above, these cleaved ends are spliced to two ends of a severed FC/APC patch cable, one of which leads to the photodiode, the other to the diode laser. This method of fiber splicing is ideal for this application, as it is quick and easy, allowing for a convenient input and removal of the tapering fiber to and from the optical circuit. Losses vary depending on fiber alignment for this splicing method, but we are only concerned with providing enough power to observe variations in fiber transmission. Once we have ensured that the splices provide sufficient power to the photodiode, the fiber is placed in the V-groove fiber holders, with the region prepared for tapering centered between them.

At this point, the hydrogen torch is lit using a butane lighter and gas flow is adjusted to ensure a steady flame about 1 cm high. This flame is then approached towards the fiber until a small (a few mm) section begins to glow, indicating that the fiber is in a molten state. Once this point has been reached, the two pulling stages move in opposite directions, each at a constant speed generally chosen to be 40 $\mu$m/s.

\begin{figure}[h!]
\includegraphics[width = \columnwidth]{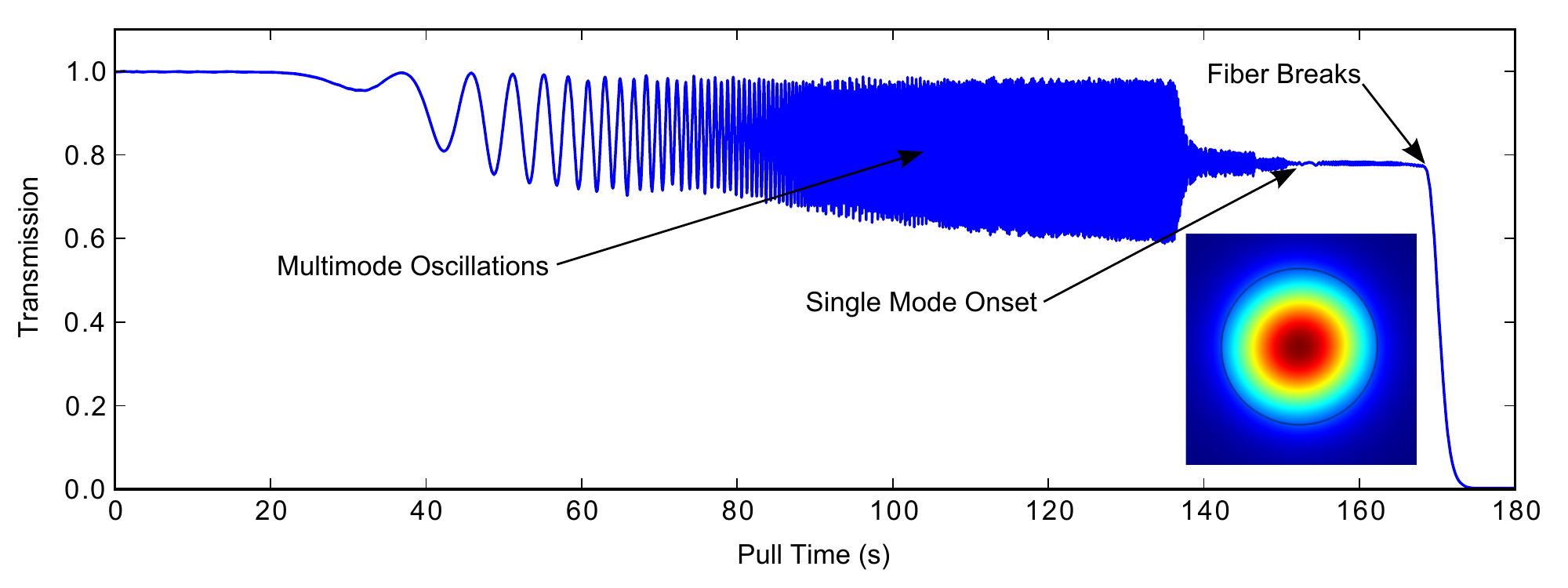}
\caption{{\bf Transmission Profile of a Tapered Fiber Pull:} Plot of transmission vs pull time for tapered fiber pull normalized to the maximum transmission through the fiber before the pull. This pull was performed using a hydrogen flame generated by the OX-00 torch tip, producing a hot-zone of $\sim$1.3 mm, at a pull speed of 40 $\mu$m/s which resulted in a final transmission efficiency of 78\%. The important regions of the transmission profile are labeled accordingly. Inset: Finite element method simulation of the time averaged energy density for the fundamental mode at 1550 nm of an air-clad tapered fiber with a diameter of 1 $\mu$m and index of refraction 1.4677.}
\label{fibtrans}
\end{figure}

\begin{figure}[h!]
\includegraphics[width = \columnwidth]{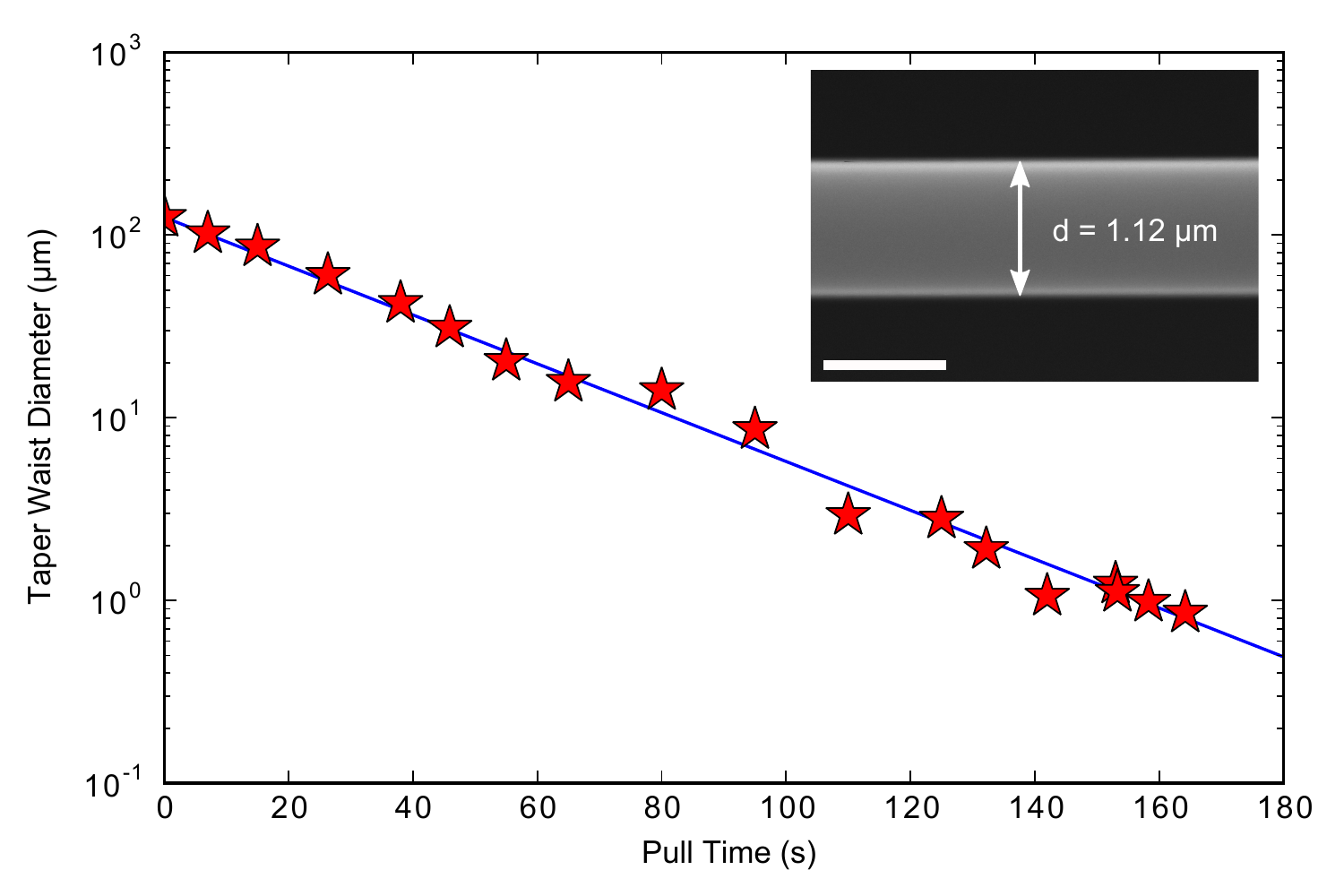}
\caption{{\bf Tapered Fiber Waist Diameter as a Function of Pull Time:} Plot of fiber waist diameter vs pull time for fibers pulled using the same parameters as described in Fig.~\ref{fibtrans}. The red stars represent experimentally measured fiber waist diameters using an SEM, while the blue line is a fit to Eq.~\ref{expd} with $L$ as the only free parameter. This fit produces a value of $L$ = 1.29 mm, in excellent agreement with the predetermined value of 1.30 mm using the single mode cutoff point. Inset: SEM image of tapered fiber waist at the single mode transition. The waist diameter is measured to be 1.12 $\mu$m. Scale bar is 1 $\mu$m.}
\label{fibdiamvstime}
\end{figure}

During each pull, the transmission through the fiber vs pull time is monitored, an example of which is presented in Fig.~\ref{fibtrans} for the OX-00 torch tip. By monitoring fiber transmission, it is possible to determine the point at which the fiber waist has become single mode. This will be indicated as a stabilization of the fiber transmission (which is evident in Fig.~\ref{fibtrans}) due to the fact that the lossy, higher order modes of the fiber have died out, leaving behind the single fundamental mode of the fiber. Using images from a scanning electron microscope (SEM - inset of Fig.~\ref{fibdiamvstime}), we experimentally measured the diameters of our fibers at the single mode transition to be $\sim$1.1 $\mu$m, consistent with the theoretically predicted diameter for an air-clad fiber with an index of 1.47 (we expect our fibers to have an index of 1.4677) at 1550 nm (see Table \ref{cutofftab}). By measuring the time required to reach this transition from a single pull, it is possible to determine a value for the hot-zone length $L$ by inverting Eq.~\ref{expd}, provided that the pull speed and initial fiber diameter are known {\it a priori}. Using this parameter, we are able to predict the fiber waist diameter for a given pull time. Note that in order for this prediction to be accurate, care must be taken to ensure that all subsequent pulls have conditions matching the orginal one in order to ensure a consistent hot-zone length. This is readily accomplished using our system. A plot of fiber waist diameter vs pull time using the apparatus described here is presented in Fig.~\ref{fibdiamvstime}, indicating excellent agreement between the hot-zone length of 1.30 mm determined using the single mode cutoff point and the fit value of 1.29 mm. This ability to predict the fiber waist diameter is useful, as it allows for fabrication of fibers whose diameters support a propagating mode that is phase matched with the resonance we are interested in, enhancing coupling of light from the tapered fiber to the optical resonator \cite{knight}.

At the point of single mode transition, the fiber waist diameter is small enough to produce the desired evanescent field required for coupling to an optical cavity, which can be seen in the inset of Fig.~\ref{fibtrans}. However, it is often advantageous to continue pulling fibers to smaller diameters, further increasing the extent of the evanescent field outside the fiber geometry, allowing for a larger range of coupling before the fiber contacts the optical resonator. It is possible to create these sub-$\mu$m diameter fibers by continuing to pull for a small amount of time ($\sim$10 s) after the single mode transition has been reached. Using the OX-00 torch tip, diameters as small as 850 nm can be achieved before the fiber breaks due to the pressure of the flowing hydrogen gas from the torch. By using the HT-3, our largest torch tip, the flame size increases, nearly doubling the hot-zone to 2.4 mm, allowing for the fabrication of tapered fibers with diameters down to 500 nm and 98\% transmission. This provides fibers with diameters small enough that they can be used as a probe of nitrogen vacancy center photoluminescence \cite{fu}, as well as allow single mode guiding of 780 nm light (Table \ref{cutofftab}), which is used in aqueous biosensing applications \cite{dantham}.

By monitoring transmission before and after the pull, it is also possible to determine the losses induced in the fiber due to the tapering process. This is important for determining the amount of power injected into the optical resonator, allowing for calculation of the number of photons confined in the optical resonator. For the OX-00 tip, a tapered fiber transmission efficiency of up to $\sim$80\% is achieved. By using the HT-3 tip, with its larger hot-zone, a more adiabatic taper transition region is created allowing us to produce fiber tapers with transmission effiencies exceeding 99\%, which is on par with state-of-the-art, ultralow loss fiber pullers \cite{ding}.

\subsection*{Fiber Dimpling Procedure}

\begin{figure}[h!]
\includegraphics[width = \columnwidth]{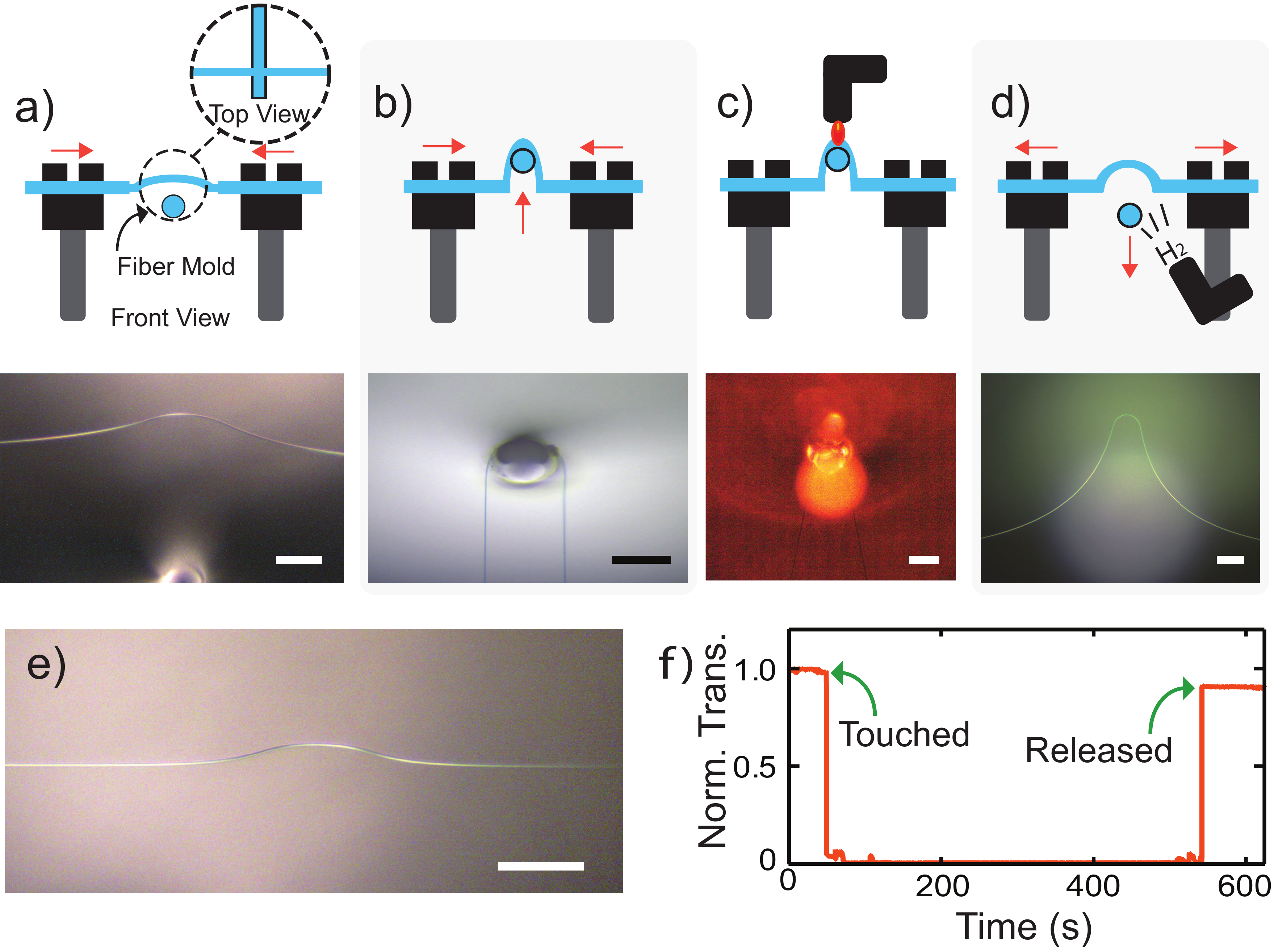}
\caption{{\bf Dimpling Procedure:} Schematics and microscope images illustrating the tapered fiber dimpling procedure. All scale bars are 125 $\mu$m. a) The fiber is detensioned slightly (by about 10 $\mu$m) by moving the fiber mounts inward, producing a small protrusion at the fiber's thinnest point. The fiber mold is then adjusted such that it is aligned with this section of the fiber. b) The fiber mold is raised approximately 5 mm allowing the tapered fiber to wrap around it. The fiber mounts are gradually moved inwards as to prevent the fiber from breaking, while still maintaining tension on the fiber mold. c) An inverted hydrogen torch with a large flame (using HT series torch tip) is approached by hand, annealing the tapered fiber into a dimpled shape. d) The fiber mold is lowered while the fiber mounts are moved outwards to restore tension to the newly formed dimple. The dimple is gently removed from the fiber mold by flowing low pressure hydrogen gas from below. e) Optical microscope image of the resulting dimpled tapered fiber using this procedure. f) Plot indicating transmission (normalized to the pre-dimpled value) through the tapered fiber before and after dimpling. Losses induced by introducing the dimple to the fiber are around 8\%.}
\label{dimple}
\end{figure}

Once a tapered fiber has been pulled, it is possible to proceed with the dimpling procedure. We begin by taping a stripped Corning SMF-28e optical fiber to the $xyz$-positioning stage located opposite the hydrogen torch, mounting it perpendicular to the tapered fiber so that it can be used as a mold in the dimpling process (see Fig.~\ref{dimple}a). The fiber mold is prepared by stripping off its acrylic coating and cleaning it with a solvent, producing a mold of 125 $\mu$m in diameter. In addition, graphite powder (SLIP Plate$^{\textregistered}$ Tube-O-Lube$^{\textregistered}$) is applied to the fiber mold to prevent it from sticking to the tapered fiber. This graphite generally burns away when introduced to the hydrogen flame during the dimple annealing process, however, using too much graphite should be avoided as it can contaminate the dimple, inducing losses. To prevent this from happening, a fiber wipe or compressed air can be used to gently remove excess graphite.

To continue, the torch is replaced by the microscope imaging system on the torch positioning gantry so that dimpling can be observed in real time. While watching with the microscope, the tapered fiber is detensioned by approximately 10 $\mu$m to reveal its thinnest point, which appears as a small bend upwards in the fiber (see Fig.~\ref{dimple}a). The stripped fiber mold is centered on this point and manually raised to touch the tapered fiber using the $z$-positioning stage. The mold fiber is then raised approximately 5 mm, while simultaneously detensioning the tapered fiber, allowing the fiber to wrap itself around the mold producing the desired dimpled shape, as shown in Fig.~\ref{dimple}b. During this process, the tapered fiber should remain tensioned tightly around the mold at all times to prevent it from twisting.

At this point, a hydrogen flame produced by the tapering torch is introduced to anneal the fiber into a dimpled shape. For this process, one of the HT series torch tips is used, producing a wide flame allowing for the increase in heat distribution required for annealing. This flame is approached to the dimple by hand, touching the mold and tapered fiber lightly (for about one second) until it glows red (see Fig.~\ref{dimple}c). The mold fiber is then slowly lowered in the same manner it was raised, this time tensioning the tapered fiber, until the mold is returned to its initial position. The dimple is then removed from the mold by using the unlit torch to flow hydrogen from below, applying a gentle pressure which releases the dimpled fiber. Typically, this process returns a dimple with minimal losses ($\sim$8\%, see Fig.~\ref{dimple}f). A microscope image of a dimpled fiber produced using this procedure is shown in Fig.~\ref{dimple}e.

\subsection*{Gluing Procedure}

\begin{figure}[h!]
\includegraphics[width = \columnwidth]{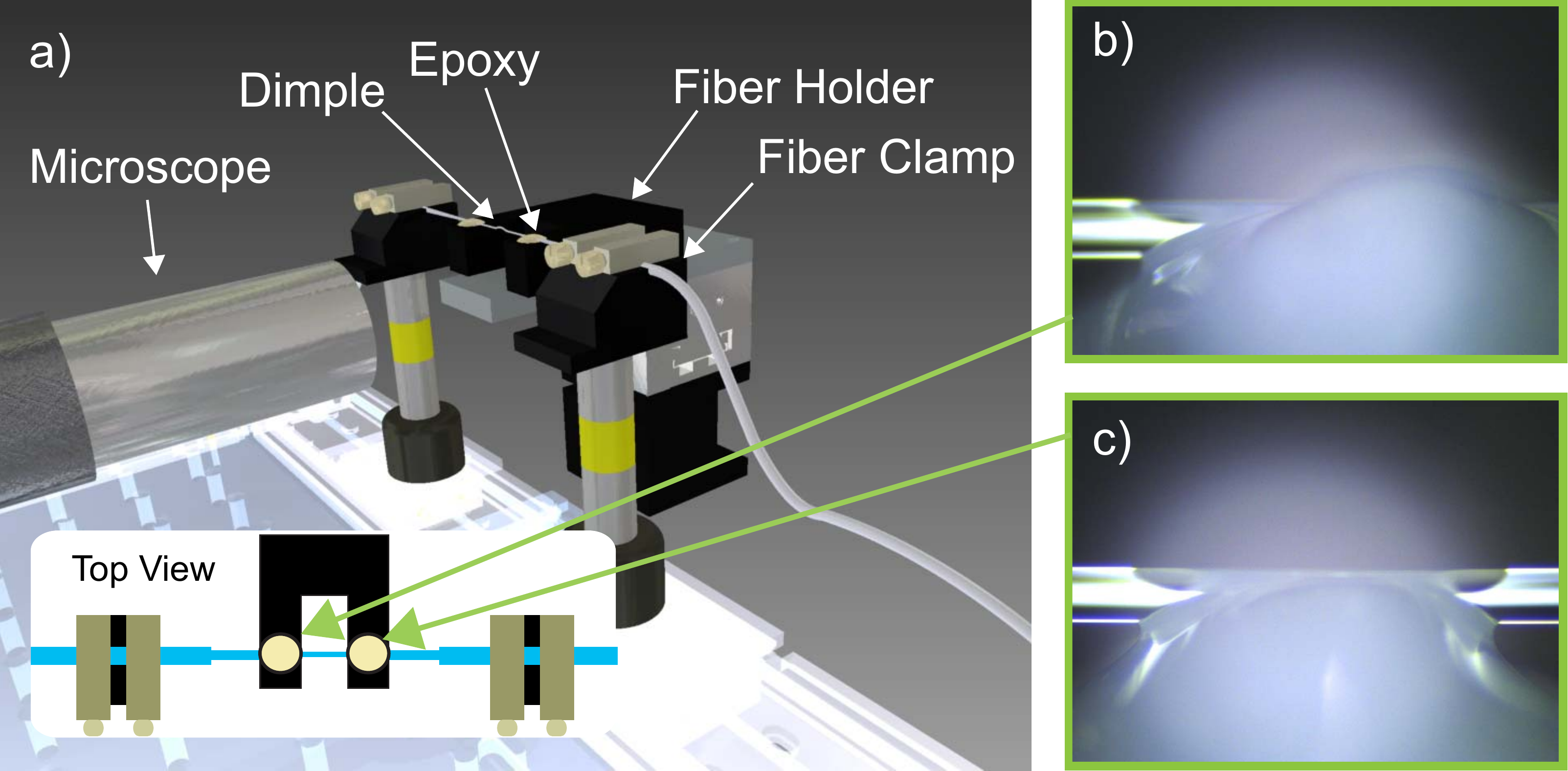}
\caption{{\bf Dimpled Fiber Gluing:} a) Labeled schematic of the fiber gluing process. Once the tapering/dimpling process is complete, two droplets of epoxy located on a fiber holder are approached to the newly created dimpled fiber using the manually adjustable stage behind the fiber mounts. This entire procedure is observed real time to ensure proper gluing using the microscope imaging system mounted on the torch positioning gantry. Inset is a top-down view of this schematic, depicting the fiber, fiber holder and fiber clamps. b), c) Microscope images taken from the mounted imaging system showing the two ends of a fully glued dimpled fiber.}
\label{gluefig}
\end{figure}

Once a dimpled tapered fiber (or other optical component created by the fiber heating system) is produced, it must be carefully attached to its holder using the gluing apparatus. To begin this process, Devcon 5 Minute$^{\textregistered}$ epoxy gel (No. 14240) is applied to both sides of the fiber holder, which can be seen in Fig.~\ref{gluefig}a. Care is taken to ensure that both droplets of epoxy are approximately the same height, ensuring that they will contact both sides of the tapered region at the same time. Once the epoxy is applied to the fiber holder, it is placed on its holding plate located on the gluing apparatus. The fiber holder is then carefully aligned beneath the fiber, ensuring that the fiber will be glued in the appropriate place. Next, the fiber holder is slowly raised using the $z$-axis of the positioning stages until the fiber has been enveloped in epoxy on both sides of the taper. This initial epoxy is then left to dry (for about 30 minutes) allowing the fiber to be rigidly held on the fiber holder, drastically increasing its durability. Once the initial epoxy dries, a second round of gluing is typically applied to the fiber, which increases the strength of the fiber's attachment to the holder.

This entire gluing process is monitored in real time using the microscope imaging system mounted on the positioning gantry, which is very helpful as we are able to definitively determine the point at which the fiber has been glued. As well, by imaging the tapered region, along with monitoring transmission down the fiber, we can determine whether or not the tapered fiber has survived the gluing process. Once the fiber has been properly glued in place, it can be transferred directly to the coupling chamber where it is fusion spliced to an existing optical circuit, allowing for injection of light into optomechanical devices.

\subsection*{Optomechanical Coupling Chambers}

\begin{figure}[h!]
\includegraphics[width = \columnwidth]{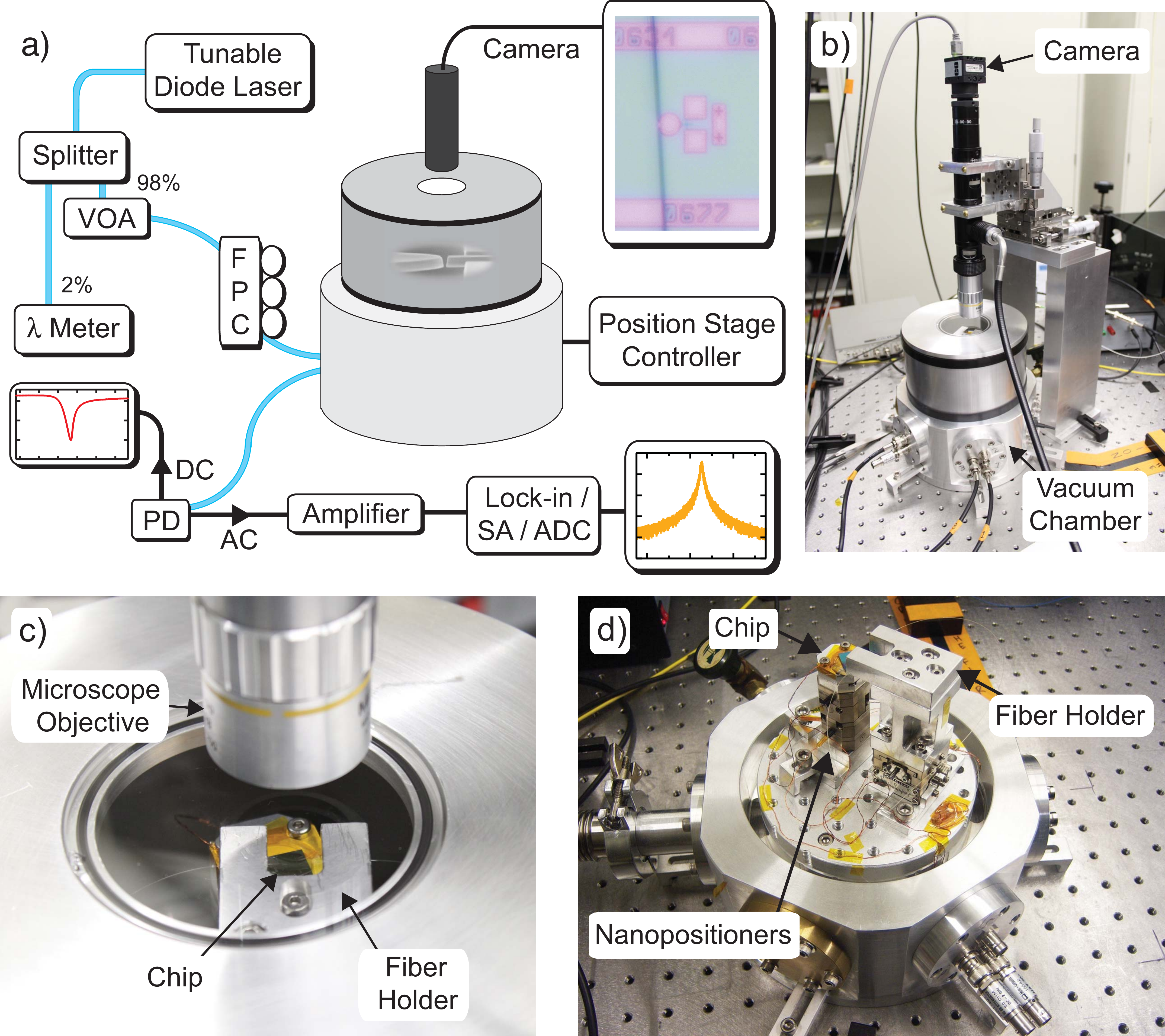}
\caption{{\bf Optomechanical Coupling Chamber:} a) Schematic of the optomechanical coupling chamber. The resonance lineshape from an optical cavity measured by monitoring the DC component of the fiber's transmission while sweeping the laser frequency across one of its resonances is shown in the bottom left in red. Likewise, mechanical resonance data measured using the spectrum analyzer (SA) is presented on the bottom right in yellow. In the top right is a picture taken with our microscope imaging system of a dimpled fiber coupled to a torsional optomechanical device. As well, an angled SEM image of the same device (not to scale) is overlayed on the chamber. FPC - fiber polarization controller, VOA - variable optical attenuator, ADC - analog-to-digital converter, PD - photodiode. b) Picture of our optomechanical coupling chamber viewed from the outside while under vacuum. c) Close-up of the optical access port, illustrating how optomechanical coupling is observed. d) Picture of the inside of the chamber containing the Attocube ANPz/x nanopositioning stack, highlighting the positioning stages and fiber holder.}
\label{optcham}
\end{figure}

Our coupling chambers, which can be seen in Fig.~\ref{optcham}, contain two separate positioning systems, with similar principles but different translation stages. In one such setup, the sample chip is placed on top of a stack of Attocube linear nanopositioning stages consisting of one ANPz101 stage mounted on top of two perpendicularly oriented ANPx101 stages. The chip is attached rigidly to a custom-machined adapter, which is fastened to the top $z$-positioning stage. This arrangement provides positioning with sub-nm precision over a total range of 5 mm. A picture of this setup can be seen in Fig.~\ref{optcham}d.

The other positioning system is built using Newport Agilis$^{\rm TM}$ AG-LS25V6 vacuum compatible, piezo driven linear stages. Two of these stages are stacked on top of each other, resulting in orthogonal $xy$-positioning, with a third mounted at a 90 degree angle for $z$-translation using an EQ3 Series angle bracket purchased from Newport. As above, the chip containing our optomechanical devices is mounted on a custom-built platform, which is attached to the vertical translation stage providing full three-axis control. These stages provide 50 nm stepping resolution over their entire travel range of 12 mm. 

Each of these systems have different strengths, with the Attocube stack providing extremely precise positioning over a relatively large range, while the Agilis stages provide a more durable, inexpensive alternative with a larger range of motion. These positioning systems are used to approach the optomechanical devices found on the sample chip to a stationary dimpled tapered fiber, which is glued to a custom-machined fiber holder. The fiber was chosen to remain fixed as it is far less stable than the devices on the chip, so its mechanical noise is reduced by anchoring it to an immobile fiber holder. 

To allow interchangability between our two chambers, this positioning system is fastened to a custom-machined circular plate, containing 1/4"-20 tapped holes in a square pattern with a spacing of 3/4". This plate is then screwed onto a homemade aluminum base with 6 ports, each of which is sealed with an O-ring and provide electrical and optical input/output for the setup, as well as allowing for pressure control inside the chamber. 

The optical input/output port consists of fiber feedthroughs, each of which allows for both an input and output fiber, channelling light to and from the tapered fiber. Each fiber is glued in place using Varian Torr Seal high vacuum epoxy, which provides the appropriate seal required for vacuum. For the Attocube setup, the electrical port houses three hermetically sealed BNC feedthroughs. The other type of electrical port, which provides input/output for the Agilis stages, is comprised of a vacuum compatible 15-pin D-type connector housed in a KF50 feedthrough flange (Accu-Glass Products - Model No. 15D-K50).

The vacuum environment provided by our coupling chambers removes airborne contaminants which can reduce the quality of the tapered fiber and optomechanical devices over time \cite{fujiwara}. It is also possible to remove such contaminants using a nitrogen purged environment \cite{borselli}, however, performing optomechanics in vacuum has the added advantage of increasing the mechanical quality factors of devices by drastically reducing viscous damping \cite{verbridge}. The vacuum pump port is comprised of a KF25 adapter connected to a turbo pump backed by a dry scroll pump. By using a completely dry pumping system, we ensure that no oil is ever backstreamed into our system. This connection is made using vibration isolating bellows, which are passed through a cement block to further prevent vibrations from the pump reaching the optical table where the chamber is located. Using this pumping system, we can achieve chamber pressures as low as 10$^{-6}$ torr. There also exists a release port, consisting of a Nupro B-4HK brass bellows-sealed valve, which allows for surrounding air to enter the system, re-establishing ambient pressure inside the chamber. All unused ports on the chamber base are covered with a blank port. This entire system is leak-checked using an Adixen ASM380 dry leak detector, ensuring it is properly sealed.

On top of the chamber base is an aluminum cylinder approximately 10 cm long and 17 cm in diameter which provides the housing for the positioning stages and tapered fiber mount. An L-shaped boot gasket (Duniway Stockroom - Part No. VBJG7) is placed on each side of the cylinder providing a leak tight seal between it and both the base and its custom-machined lid. Optical access through the lid is provided by a 75 mm diameter optical flat glass window (Edmund Optics 1/4-Wave N-BK7 - Stock No. \#62-606), which lays flush against an O-ring located in a recessed portion of the lid when the chamber is under vacuum. This window is located directly above the tapered fiber holder and positioning stages, which allows for real time monitoring of the optomechanical system (see Fig.~\ref{optcham}c). It is therefore possible to view the tapered fiber and on-chip devices while attempting to couple between them, which is important for this process. The fiber and chip are imaged using the exact same imaging system described above for the tapered fiber puller. This is made possible by the fact that this microscope is oriented using a three dimensional arrangement of manually positioned New Focus 9063-COM gothic-arch translation stages mounted on a two legged custom-built stand with identical mounting plate to that used for manual $z$-positioning of the hydrogen torch in the tapered fiber puller, allowing interchangibility of the micrscope between the two systems. The Mitutoyo No. 906912 micrometers used to manipulate this positioning system, provide a 25 mm travel range with 10 $\mu$m resolution. This resolution is more than enough to view our devices in the $xy$-plane of the chip, as well as provide excellent focusing for our imaging setup.

\subsection*{On-Chip Optomechanical Devices}

\begin{figure}[h!]
\includegraphics[width = \columnwidth]{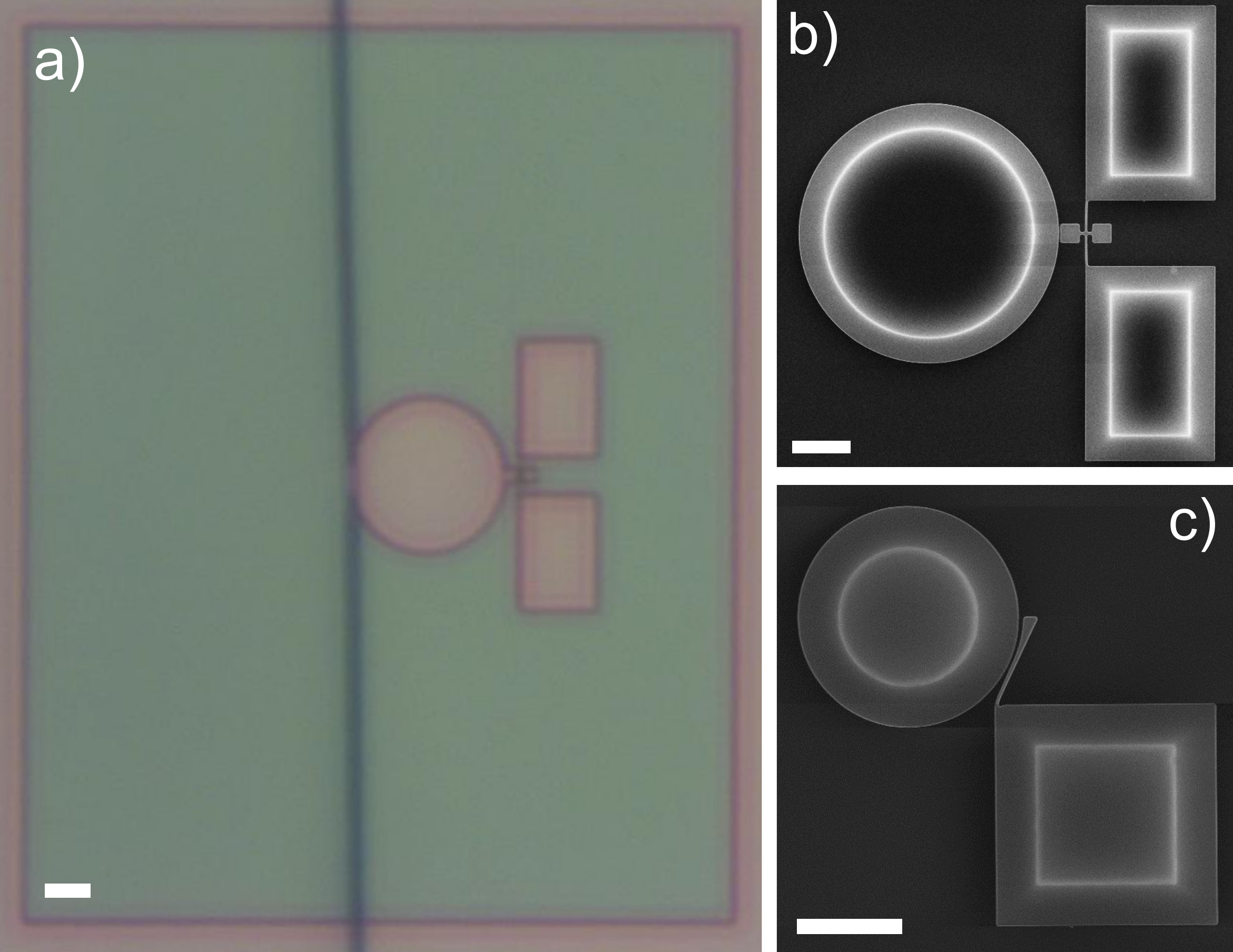}
\caption{{\bf Optomechanical Devices:}  a) Optical micrograph of an on-chip torsional optomechanical device taken with the imaging system used to observe the optomechanical chamber. The green indicates etched regions, while the remaining device layer is shown in pink. The large etched region provides access for the dimple to couple to the microdisk. On the right are SEM images of b) a torsional mechanical device and c) a microcantilever, both of which are side-coupled to a microdisk WGM resonator. Note that the cantilever follows the curvature of the disk to enhance optomechanical coupling. All scale bars are 5 $\mu$m.}
\label{devfig}
\end{figure}

Due to the small feature sizes required for our optomechanical devices, we have chosen to use foundry-based nanofabrication, which
provides high throughput of devices with a minimum feature size of 100 nm using top-down DUV photolithography \cite{selvaraja}. Each optomechanical device consists of an optical microdisk side-coupled to a mechanical nano/micro-resonator, such as a torsion paddle or a cantilever, as can be seen in Fig.~\ref{devfig}. Each device is centered in a large etched area (approximately 100 $\mu$m $\times$ 50 $\mu$m), which provides ample room for coupling using our dimpled tapered fiber method. The mask for these devices is designed using custom-programmed Python scripts utilizing the gdspy module, which generates a GDSII file containing our chip layout. This allows us to iterate through a number of device specifications, such as coupling gap, disk radius and mechanical resonator dimensions, providing a large parameter space in which we can explore different optomechanical regimes. These design files are then submitted through Canadian Microelectronic Corporation (CMC) Microsystems to the Interuniversity Microelectronics Center (IMEC) located in Leuven, Belgium. It is here our devices are fabricated on an 8 inch silicon-on-insulator (SOI) wafer, which consists of a 220 nm thick layer of single crystal silicon supported by a 2 $\mu$m layer of silicon dioxide. The single crystal silicon device layer is ideal for optomechanical devices, as it has negligible absorption in the telecom band around 1550 nm and a high index of refraction ($n \approx 3.42$), enhancing the mechanical resonator's perturbation of the optical cavity's evanescent field. Our devices are patterned onto this wafer using an excimer laser (193 nm - 248 nm) and high-definition photomasks derived from our design files. These patterned wafers are then etched using either a standard or high dose recipe, producing optomechanical devices in the silicon layer which are held rigidly in place by the oxide buffer. This, along with a protective resist coating the entire wafer, help to prevent damage to the devices during transit.

After we receive these wafers, a number of post-processing procedures must be performed in order to prepare our devices for measuring. We begin by dicing the 8 inch wafer wafer into 1 cm $\times$ 1 cm chips using a diamond saw. Each chip is then ultrasonically cleaned with acetone and rinsed with isopropyl alcohol to remove the protective coating. Once the wafer has been diced and cleaned, a buffered oxide etch (BOE) is used to selectively remove the sacrificial oxide layer beneath our mechanical devices, which releases them, allowing them to oscillate freely. It is important to note that since BOE is a wet etch, we must ensure that our devices are dried using either a critical point drier or ultralight solvents, such as n-Pentane (C$_5$H$_{12}$), to prevent stiction. Once a chip has been etched and dried, it is ready to be placed in the chamber for measuring.

\subsection*{Coupling Procedure}

Coupling to our optomechanical devices begins by locating the dimple of the tapered fiber using the imaging apparatus. This is done by searching for the portion of the fiber that is in focus at the lowest point (due to the fact that the dimple protrudes away from the rest of the fiber). Once the dimple is found, the nanopositioning stages are used to align the desired optomechanical device such that the lowest point of the dimple is able to couple light into the optical resonator.

The precision of our nanopositioning stack allows for two methods by which we can couple light into the modes of our optical resonators. We can either bring the fiber close enough to the device such that the cavity's optical modes are excited by the fiber's evanescent field or we can simply touch the fiber to the optical resonator. Hovering has the advantage that the excited optical modes are less perturbed by the fiber's presence, resulting in reduced losses. However, by touching the fiber to the resonator, the mechanical instability of the fiber is further reduced. As well, by using this coupling method, it is possible to excite a larger number of optical modes, some of which have higher $Q$s and larger optomechanical coupling to the mechanical resonator.

\subsection*{Data Acquisition: Side-of-Fringe and Homodyne Detection}

\begin{figure}[h!]
\includegraphics[width = \columnwidth]{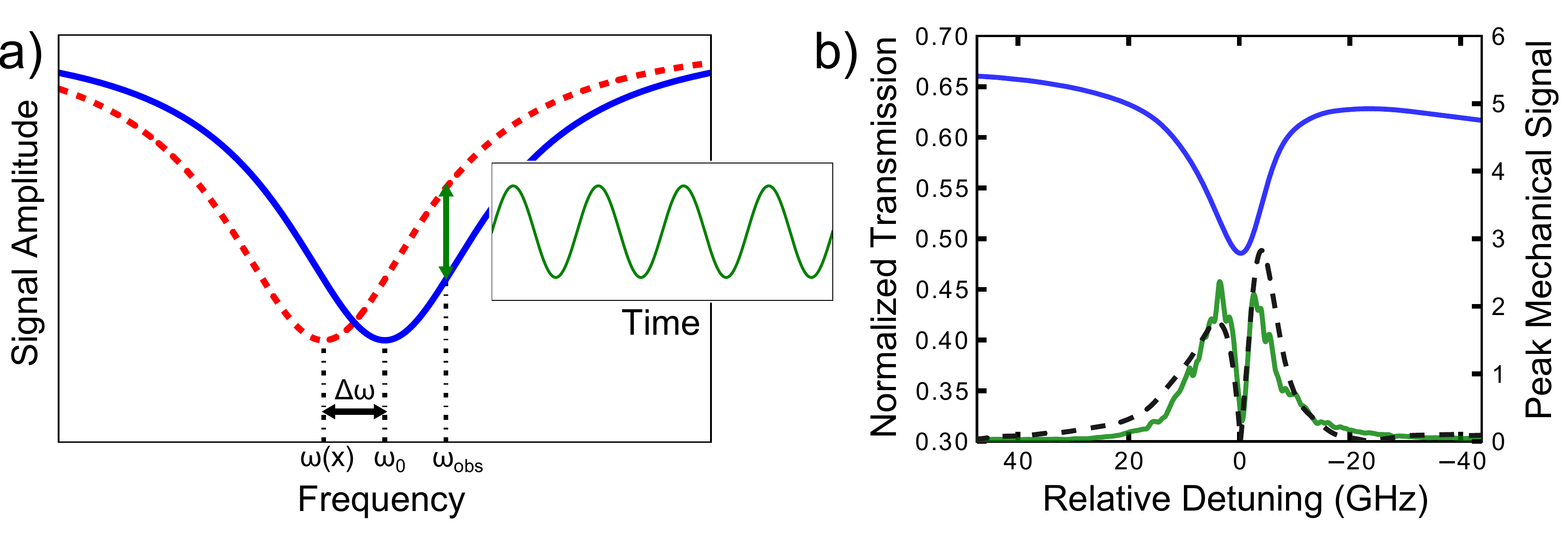}
\caption{{\bf Optomechanical Data Acquisition:} a) Schematic of the mechanism by which ``tuned-to-the-slope'' cavity optomechanical detection is performed. We begin with an unperturbed optical cavity, with a resonance lineshape (solid, blue) centered at $\omega_0$. The periodic displacement $x$ of the mechanical resonator shifts this resonance by an amount $\Delta \omega$, producing a new trace (dashed, red) centered at $\omega(x)$. By tuning our laser to the observation frequency $\omega_{\rm obs}$, the optical transmission (inset, green) oscillates in time at the mechanical device's resonance frequency. This signal is maximized by choosing $\omega_{\rm obs}$ to be the value at which the optical resonance has maximum slope. Note that the schematic is exaggerated to better illustrate this effect. b) Plot of optomechanical data taken for a microcantilever side-coupled to an optical microdisk, as seen in Fig.~\ref{devfig}c. The solid blue trace is produced by scanning our laser frequency around an optical cavity resonance and monitoring DC transmission through the tapered fiber coupled to the device. While performing this scan, we simulataneously monitor the AC transmission signal through the fiber at the mechanical device's resonance frequency, using a lock-in amplifier. By virtue of our optomechanical transduction scheme, this signal represents the peak value of the mechanical motion and is plotted in green. The dashed line is a numerical derivative of the optical trace normalized such that is has the same peak value as the green data, which indicates that mechanical signal is optimized at the points of highest slope in the optical resonance.}
\label{coupfig}
\end{figure}

Once we have coupled light into our optical resonators, we begin measuring our devices' mechanical motion using amplitude sensitive measurements in the ``tuned-to-the-slope'' regime \cite{braginsky}, which is illustrated in Fig.~\ref{coupfig}a. In this detection scheme, the mechanical device's motion causes the cavity resonance to shift, which is transduced by the slope of this lineshape into AC transmission fluctuations in the fiber, occuring at the mechanical resonance frequency. Therefore, by tuning our laser wavelength to the maximal slope of our optical cavity resonance, we provide optimal optomechanical transduction efficiency, as can be seen in Fig.~\ref{coupfig}b. In general, this resonant enhancement scales with the system's optical quality factor which increases the slope of the resonance, however, this is convolved with other effects such as an optical mode's volume and overlap with mechanical motion \cite{kim}. Using this method, we have probed devices with an angular resolution of 4 nrad/$\sqrt{\rm Hz}$ corresponding to a torque transduction on the level of 4$\times$10$^{-20}$ N$\cdot$m/$\sqrt{\rm Hz}$ \cite{kim}, as well as displacement noise floors of 2 fm/$\sqrt{\rm Hz}$ and force sensitivity of 132 aN/$\sqrt{\rm Hz}$ \cite{doolin}.

It is also possible to perform phase-sensitive measurements on our devices in the ``tuned-to-the-peak'' regime using a balanced optical homodyne detection system \cite{yuen}, which can be used for quadrature and entanglement measurents \cite{laurat}.  This method of detection also has a number of advantages, including cancellation of laser noise \cite{safavi} and the ability to lock the laser to the bottom of the optical resonance \cite{schliesser}. Furthermore, since the laser's detuning from the cavity resonance is zero, a maximum number of photons are coupled into the cavity, which enhances the system's optomechancal coupling.

After setting up one of these detection schemes, the AC transmission signal through the fiber is sent to a spectrum analyzer (SA), which outputs its frequency power spectrum. For an optomechanical signal, this includes the power spectral density (PSD) of the mechanical motion \cite{hauer}. 

Alternatively, a time series measurement of the voltage signal taken with an analog-to-digital converter (ADC) can be digitally analyzed to determine its spectral components. In our system, this is performed using a digital lock-in amplifier (Zurich Instruments HF2LI), which is a specialized ADC that first mixes an input voltage signal with a reference frequency, $\omega_{\rm ref}$, shifting the frequencies of the input signal by $\pm \omega_{\rm ref}$. Therefore the frequency information around the reference frequency of the input signal is now the low frequency component of the mixed signal. This has the advantage of requiring a data collection rate proportional to the bandwidth of the signal measurement, as opposed to a data collection rate proportional to the maximum frequency component.  For example, when bandwidth of only 100 kHz centered at 10 MHz contains important spectral information, a data collection rate of $\sim 10^5$ samples per second (SPS) can be used as opposed to a rate of $\sim 10^7$ SPS, a reduction of about 100 times the data needed to acquire the signal of interest.

After mixing, the lock-in applies a low-pass filter to reduce noise contributions from unwanted frequencies outside the measurement bandwidth,  thus the time series output of a lock-in amplifier is the result of a convolution of the lock-in amplifier's filter response with the demodulated input signal, \textit{i.e.}

\begin{equation}
Z(t) = X(t) + i Y(t) = \{H(t) * e^{i \omega_{\rm ref} t} V(t)\} (t),
\end{equation}

\noindent where $X(t)$ and $Y(t)$ are the two outputs of a dual-phase lock-in amplifier,  $H(t)$ is the impulse response of the lock-in amplifier's filter, and $V(t)$ is the input signal to the lock-in.  Fourier transforming the output elucidates the convolution, giving

\begin{equation}
Z(\omega) = H(\omega)  V(\omega - \omega_{\rm ref}).
\end{equation}

\noindent Thus the spectrum of the lock-in amplifier's output is the spectrum of the input voltage translated in frequency by the reference frequency and enveloped by the lock-in amplifier's filter.  The power spectrum can then be estimated by taking $S_{\rm ZZ} = |Z(\omega)|^2$, or done in practise by using a PSD estimation algorithm such as Bartlett's method \cite{bartlett}, giving

\begin{equation}
S_{\rm ZZ} (\omega) = |H(\omega)|^2  S_{\rm VV} (\omega - \omega_{\rm ref}).
\end{equation}

\noindent This method of data acquisition allows for real time optimization of optomechanical transduction in our devices, which facilitates sensitive probing of their mechanical motion, allowing for precise measurements of physical quantities, such as forces \cite{doolin} and torques \cite{kim}.

\section*{Conclusion}

This article presents a method by which high efficiency optical coupling is achieved between a dimpled tapered fiber and nanofabricated on-chip optomechanical devices. By using a custom-built automated heat-and-pull fiber puller, it is possible to consistently produce tapered fibers of a predetermined diameter, which is often chosen to be less than the single mode cutoff diameter, providing ample evanescent field for coupling. Dimpling this tapered fiber, using a well-defined procedure, allows for production of an excellent localized probe of planar on-chip devices. Attaching this fiber to a robust holder permits it to be tranferred to special coupling chambers. In these chambers, optomechanical coupling is performed in an optical access vacuum environment and is mediated by high precision, nanopositioning stages. By using the amplitude sensitive ``tuned-to-the-slope'' detection scheme, angular resolution of 4 nrad/$\sqrt{\rm Hz}$ \cite{kim} and displacement transduction of 2 fm/$\sqrt{\rm Hz}$ \cite{doolin} have already been demonstrated. It is anticipated that using this technique for optomechanical coupling, we will be able to continue to measure increasingly sensitive devices, approaching the measurement limits imposed by quantum mechanics \cite{braginsky}.

\section*{Competing interests}
  The authors declare that they have no competing interests.

\section*{Author's contributions}

BDH, PHK and JPD conceived and designed the experiment. BDH, PHK and AJRM constructed the experiment. PHK performed nanofabrication post-processing for the on-chip devices. BDH, PHK, CD, AJRM and HR collected and analyzed data. BDH and AJRM performed theoretical calculations regarding the single mode cutoff diameter for tapered fibers. CD and PHK developed and optimized the procedure for producing dimpled tapered fibers. CD and HR created software for data taking and system manipulation. BDH, PHK, CD and JPD drafted the manuscript. All authors have read, approved and provided critical revisions for the final manuscript.

\section*{Acknowledgements}
  The authors would like to thank Prof. Paul Barclay for numerous helpful suggestions and insight into both the theoretical and practical applications of optomechanics. We would also like to thank Don Mullin, Devon Bizuk and Greg Popowich for technical assistance. This work was supported by the University of Alberta, Faculty of Science; the Natural Sciences and Engineering Research Council of Canada; Alberta Innovates Technology Futures; the Canada Foundation for Innovation; and the Alfred P. Sloan Foundation.

\bibliography{EPJ_OnChipOptomech}

\section*{Figures}
 
\section*{Tables}

\begin{table}[h!]
\begin{tabular}{ ccccc }
\hline
\multirow{2}{*}{$\lambda~({\rm nm})$} & \multirow{2}{*}{$n_{co}$} & \multirow{2}{*}{$n_{cl}$} & \multicolumn{2}{c}{$d_c~({\rm nm})$} \\  
 & & & WGA & Numerical \\ \hline
\multirow{2}{*}{637} & 1.47 & 1.00 & 452.6  & 452.6 \\
 & 1.47 & 1.33 & 778.8 & 778.9 \\ \hline
\multirow{2}{*}{780} & 1.47 & 1.00 & 554.1  & 554.2 \\
 & 1.47 & 1.33 & 953.6 & 953.7 \\ \hline
\multirow{2}{*}{1310} & 1.47 & 1.00 & 930.7  & 930.7 \\
 & 1.47 & 1.33 & 1601.6 & 1601.7 \\ \hline
\multirow{2}{*}{1550} & 1.47 & 1.00 & 1101.2  & 1101.2 \\
 & 1.47 & 1.33 & 1895.0 & 1895.1 \\ \hline
\hline
\end{tabular}
\caption{Single mode cutoff diameter calculated using both the WGA approximation and numerical calculations for a green light (637 nm) observed in nitrogen vacancy photoluminescence \cite{fu} and near infrared light (780 nm) used in aqueous biosensing \cite{dantham}, as well as the dispersionless and low attenuation telecom wavelengths of 1310 nm and 1550 nm. All calculations are performed for both air-clad ($n_{cl}=1.0$) and water-clad ($n_{cl}=1.33$) environments. The appropriate number of digits are retained to show the difference in WGA and numerical calculations.}
\label{cutofftab}
\end{table}

\end{document}